# Accessibility and Inclusiveness of New Information and Communication Technologies for Disabled Users and Content Creators in the Metaverse

**Authors:** Dr Petar Radanliev[1*], Professor David De Roure[1], Dr Peter Novitzky[2], Dr Ivo Sluganovic[3]

[*]**Corresponding author:** Petar Radanliev petar.radanliev@oerc.ox.ac.uk

**Addresses:** [1]Oxford e-Research Centre, Department of Engineering Science, University of Oxford; [2]UCL Faculty of Engineering Science — STEaPP; [3]Department of Computer Science, University of Oxford

**Funding:** This work has been supported by PETRAS National Centre of Excellence for IoT Systems Cybersecurity, which has been funded by the UK EPSRC [under grant number EP/S035362/1], the Software Sustainability Institute [grant number: EP/S021779/1], and by the Cisco Research Centre [grant number CG1525381].

**Acknowledgments:** Eternal gratitude to the Fulbright Visiting Scholar Programme.

**Conflicts of Interest:** The authors declare no conflict of interest.

**Availability of data and materials:** all data and materials are included in the article.

**Authors contributions:** Dr Petar Radanliev and Professor David De Roure (funding acquisition, data collection and analysis, writing, review, submission, correspondence), Dr Peter Novitzky and Dr Ivo Sluganovic (review, feedback, correction, discussion, and visualizations).

**Abstract:**

(1) Purpose: The purpose of this journal article is to present a comprehensive review of current information technologies and their potential to enable and empower physically disabled users and content creators within the evolving concept of the Metaverse. It aims to critically reassess existing Metaverse concepts and propose a novel framework for purposeful design that emphasises inclusiveness for physically disabled artists. The article also highlights the need for standards and regulations governing the inclusion of people with disabilities in Metaverse projects.

(2) Materials and Methods: The study utilises a review-based approach to examine current information technologies and their relevance to the inclusion of physically disabled individuals in the Metaverse. It involves analysing existing Metaverse concepts, exploring emerging information technologies such as Virtual and Augmented Reality, and the Internet of Things. The design framework proposed in the article is based on the active involvement of disabled creatives in the development of solutions for inclusivity.

(3) Results: The review reveals that despite the proliferation of Blockchain Metaverse projects, the inclusion of physically disabled individuals in the Metaverse remains distant, with limited standards and regulations in place. However, the article proposes a concept of the Metaverse that leverages emerging technologies, such as Virtual and Augmented Reality, and the Internet of Things, to enable greater engagement of disabled creatives. This approach aims to enhance inclusiveness in the Metaverse landscape.

(4) Conclusions: Based on the findings, the paper concludes that the active involvement of physically disabled individuals in the design and development of Metaverse platforms is crucial for promoting inclusivity. The proposed framework for accessibility and inclusiveness in Virtual, Augmented, and Mixed realities of decentralised Metaverses provides a basis for the meaningful participation of disabled creatives. The article emphasises the importance of addressing the mechanisms for art production by individuals with disabilities in the emerging Metaverse landscape. Additionally, it highlights the need for further research and collaboration to establish standards and regulations that facilitate the inclusion of physically disabled individuals in Metaverse projects.

## 1. Introduction

The Metaverse is a combination of virtual worlds, augmented reality, life logging, mirror worlds and the Internet [1]. The Metaverse has existed for decades without any significant push towards mass adoption. However, the hardware and software technologies are close to reaching a 'tipping point' that would enable large-scale participation [2]. The use of virtual and augmented reality (VR/AR) has become more affordable. However, while it is true that the use of virtual and augmented reality (VR/AR) has become more affordable, it is important to consider the existing digital divides, particularly for people with disabilities. These divides refer to disparities in technology access and internet connectivity among different socioeconomic groups. Despite the affordability of VR/AR devices, there are still barriers that prevent equitable access to these technologies, especially for individuals with disabilities. It is crucial to address these gaps and ensure that efforts to make VR/AR affordable are accompanied by initiatives to bridge the accessibility divide for people with disabilities, allowing them to fully benefit from these technologies.

The ability to connect via a mobile device and pay using a virtual currency has enabled access for users previously without traditional finance infrastructure (mostly in developing countries), which lead to profits increase and overall expansion. Another reason for the current mass adoption is that *'the current Metaverse is based on the social value of Generation Z that online and offline selves are not different'* [3]. This definition is different from the 1990s definition of the Metaverse based on 'Second Life' multimedia platform, because the VR/AR technology has changed into a more immersive experience than 3D game world on a 2D screen.

The future Metaverse is based on interconnected virtual worlds, cyber-physical Internet-of-Things (IoT), artificial intelligence (AI), Blockchains, and other technologies, engaged with social, economic, and political activities. It is expected to surpass the boundaries of the current physical and digital world, including the research on healthcare and entertainment [4]. The corelations between these two industries (healthcare and entertainment) was emphasised during the Covid-19 pandemic. Recent study correlates the Metaverse with healthcare [4], and as Covid-19 slowly becomes a thing of the past, these findings should be used to assist, amongst others, artists with a variety of healthcare problems (e.g. disabled musicians, painters, dancers)—in this paper referred to as 'disabled artists' ad the overarching term for 'disabled creatives' and 'disabled users'. The review is focused on future technologies that can create positive experiences, with the aim of developing '*a set of strategies to help those who want to engage with the technology futures*' [5]. The focus on Arts is because it is one of the sectors that was hit the worst during the Covid-19 pandemic (e.g., the creative and performing arts).

Almost all activities associated with creative and performing arts (e.g., dance, music, opera, theatre) were cancelled for a prolonged period. This created multiple challenges for people employed in the sector, for the entertainment industry, and for the communities and services that depend on creative and performing art entertainment. At the same time, Covid-19 increased digital adoption and normalised a variety of online engagements, triggering strong interest in the Metaverse. The human society has never been closer to reaching the human-machine symbiosis, which is defined as *'a multivariate perspective for physically coupled human-machine systems'* [6].

## 1.1. Current Issues with Inclusiveness in the Metaverse

This study reviews two current issues associated with the Metaverse inclusiveness of physically disabled creatives. First, the increased mass adoption, without appropriate attention and engagement with physically disabled creatives. Second, the lack of financial profit from interoperability, that presents an opportunity to investigate the key stages of the Metaverse development and make recommendations on inclusiveness as a crucial concept of the Metaverse. In other words, if a specific accessibility solution is developed on a specific Metaverse technology, there is no incentive for other Metaverses to adapt to a non-native technology that benefits the competing Metaverse project. Interoperability could be the key for adoption of technologies for accessibility and inclusiveness of disabled creatives in the Metaverse.

The Metaverse is a virtual world in the cyberspace that continues to exist when one is offline. This virtual world integrates the cyber-physical, and can be accessed via VR sets, PCs, and even smart phones. Since the beginning of the Covid-19 pandemic, the digitalisation of our society has increased at unprecedented rate. This resulted in an increased adoption of new technologies, creating the booming Metaverse, not only in VRs, but also in our everyday lives (e.g., working from home in a Metaverse virtual store, or virtual concert). In times of emergency little attention has been placed on inclusiveness and enabling physically disabled users to engage with the Metaverse. One of the risks from the increased adoption of new technologies without considering the needs of disabled users, is the exclusion of physically disabled creatives from one of the key new technologies that can help disabled people to operate in our society (e.g., work and socialise in hybrid physical, virtual, and augmented realities).

Moreover, the existing Metaverse concept defines a new interoperable digital economy, centred around portable digital assets or non-fungible tokens (NFTs). The existing Metaverse has gained significant attention in recent years, and Covid-19 has normalised a variety of online engagements. Adoption is, however, limited to artists and organisations with resources to recruit technical experts. Further complexity emerges because interoperability is not profitable or desirable for individual projects; and because the computers at present lack the required optimised and sustainable computing power. This outlines a major limitation of the current Metaverse concept and describes why the current Metaverse is a very limited version of the future potential of the Metaverse concept. This presents an opportunity to investigate in real-time the crucial stages of the development and the evolution of the current Metaverse into a new and more inclusive concept of the Metaverse.

## 1.2. Disability and Inclusiveness in the Metaverse

This review study begins with **two hypotheses**. The **first hypothesis** motivating this study is 'the Metaverse can be accessible (usable) and inclusive of people with physical and cognitive disabilities.' **Second hypothesis** is 'physically disabled artists can perform (create) in the Metaverse'. The review initiates with these two hypotheses as the guiding principles for encouraging disabled creatives to engage with the Metaverse. The review is designed to inspire creativity in disabled artists, encouraging creative and performing artists to engage with the Metaverse. The Metaverse idea promotes a form of digital society, where the *'concepts of race, gender, and even physical disability would be weakened'* [7]. The review is structured around two **key review categories**. **First** is to review all work on designing a purpose-driven VR/AR specifically targeted at enabling physically disabled artists to engage with the Metaverse. The focus in this category is placed on purposeful design, and meaningful consumer experiences [8]. **Second** is to review all work on determining if the physical and cognitive demand, the resources, and cost requirement are different (lower, similar, or higher) than in physical reality. A recent study on the challenges of entering the Metaverse determined that resources and cost of operating in the Metaverse are *'different and higher than physical reality,'* while the physical and cognitive demand frustration, performance, effort, physical, mental, and temporal demand was *'significantly associated'* with mental demand [9].

### 1.3. Review on Supporting Physically disabled Creatives to Engage the Metaverse

The **first phase** of the review study initiates with searching and analysing records from the Web of Science data reciprocity to identify how physically disabled creatives (artists) can be empowered to use the Metaverse. The first data reciprocity used is the Web of Science Core Collection[1], and secondary data is collected and analysed with the Web of Science analyse result tool[2]. Considering that the proposed topic has not been studied extensively and considering that the Metaverse is still a new technology, the secondary data from existing databases is complemented with collecting additional secondary data from open-source intelligence (OSINT).

The **second phase** of the review applies interviewing methods to collect primary data on how Metaverse technologies (e.g., VR/AR) can be re-designed to be accessible and inclusive of disabled artists needs and requirements. This second phase identifies the tools and technologies that empower users with physical and cognitive disabilities (e.g., enabling Metaverses to allow users to operate haptic gloves to touch and feel art masterpieces on display).

The **third phase** of the review is to identify emerging novel opportunities for disabled artists (e.g., getting involved in social and community events) and solutions for removing physical and cognitive barriers. This phase consults the PETRAS National Centre of Excellence for IoT Systems Cybersecurity[3], to identify the values, risks, and opportunities, for disabled creatives from an ecosystem of millions and billions of connected IoT or edge devices. The low cost, combined with the high volume, makes IoT devices valuable for building upon real world environments inside the Metaverse.

The **fourth phase** of the review is focused on re-designing the concept of assistive communication tools in the Metaverse, with the support of disabled artist (designed by disabled artists—for disabled artists), to ensure that assistive technologies can be embedded in and be capable of improving the accessibility of Metaverse wearables. This phase of the review will include a combination of experts in accessibility and disabled creatives and guided by a set of technologies used in the current version of the Metaverse.

The **final phase** of the review discusses the development of the parameters of a new framework, and a diagram that comprehensively displays our findings and draw new conclusions.

## 2. Research Methodology

The article is structured in two review categories with specific research tasks that enable the design of a (1) purpose-driven 'concept Metaverse,' specifically targeted at enabling disabled artists to engage with the Metaverse; and (2) determine how the physical and cognitive demand, the resources, and cost requirements can be lowered for disabled artists (comparing to physical reality). Testing of the findings is conducted with and includes experts in the field and disabled creatives. Output is structured in a new framework for purposeful design and meaningful experiences for engagement and integration of disabled creatives in the Metaverse. The proposed framework is anticipating a future need in our society to improve the functionality and enhance inclusiveness of the Metaverse for physically disabled creatives. In addition, the active engagement with physically disabled artists benefits industry with recommendations for improved products and services for disabled users and content creators, and society in general, by promoting empowerment, accessibility, and engagement with the virtual societies.

---

[1] https://www.webofscience.com/wos/woscc/basic-search

[2] https://www.webofscience.com/wos/woscc/analyze-results/baea54c2-db2c-4f9b-bf6a-c19c2d7a3984-5add94b5

[3] https://petras-iot.org/

The research methodology applies open science practices, and the data collected via review and interviews and is categorised on a) meaningful experiences, b) physical and cognitive demand, c) resources, and d) cost requirements for physically disabled creatives engaging in the Metaverse. The data categories are structured with grounded theory (GT) approach and used for constructing the new framework for facilitating and encouraging physically disabled artists to enter the Metaverse, and for deriving recommendations for creating industry standards for inclusiveness in the Metaverse. The primary data is anonymised before being made publicly available in open access (following the University of Oxford ethical procedures), for reusability and replication of results. Information and data are shared with the University of Oxford Bodleian Library, and the new framework is created in accordance with the University of Oxford regulations on ethics and data protection. Other researchers will be able to use our data and the research results are published in open access (open access fees are paid by the University of Oxford) to promote replication. This approach to open science enables further feedback from experts in this field and physically disabled creative across the world. The data management plan is developed following the guidance from the University of Oxford and the FAIR guiding principles for scientific data management and stewardship [10].

## 3. Accessibility of the Metaverse

The Metaverse can be defined as a convergence of many different cyber-physical worlds, creating virtual communities for socialising, working, relaxing, playing games, etc [11]. This convergence has been referred to as *'MetaSocieties'* and is expected to *'run in parallel'* where *'any human, enterprise, and city in the real societies will have corresponding virtual human, virtual enterprise, and virtual city, respectively, in the MetaSocieties'* [11]. Many commercial organisations (e.g., Microsoft, Nvidia, Meta) and consumer brands (e.g., Gucci, Coca-Cola, Adidas) have engaged with the Metaverse. Their advertising can be replicated and adapted by smaller companies, and two recently stated solutions for this are grounded on a *'methodological framework'* for *'conceptualization'* [12]. The value of this methodological solution needs to be reviewed and examined for advertising of less know creative and performing artists, with specific focus on physically disabled artists, and/or new and emerging artists.

### 3.1. Technical Requirements for Physically disabled Creatives to Engage the Metaverse

Large corporations and famous artists (e.g., Ariana Grande[4]) successfully secured resources and know-how to expand operations in the Metaverse. New and emerging artists lack either the technical know-how, or established followers to engage successfully at the same speed. Hence, entering the Metaverse can be a daunting task. However, one recent study on university students concluded that *'perceptions to use Metaverse were significantly associated with their innovativeness, which is, in turn, influenced by perceived ease of use and perceived usefulness'* [13]. To evaluate if *'innovativeness'* and the *'perceived ease of use'* is a factor that could help physically disabled artists enter the Metaverse and discover its *'usefulness'*, the case study research method [14] is used in combination with GT [15] to categorise the emerging data.

The Metaverse idea is not new, it originated in 1992 [16], but the modern version combines several new technologies and has many more users, and strong supporting economies. The blockchain technology has enabled creators and artists to transact and own good that can be monetised (e.g., blockchain technologies and NFTs). This review investigated how creative and performing artists—with a focus on physically disabled artists, can apply these new technologies for transacting, owning NFTs to store and exchange value, and the impact of disabled artists participating in governance of these projects/platforms. The review investigated how the Metaverse inspires creativity in physically disabled creatives, users, and content creators and categorises the values of decentralised ownership economy for disabled artists.

---

[4] https://www.bbc.co.uk/news/av/technology-58146042

## 4. Framework for Enabling and Encouraging Physically disabled Creatives in the Metaverse

The new framework builds on current and past projects (i.e., AI StyleGAN[5]; Tilt Brush[6]; DeepArt.io[7]; Sougwen Chung[8]; EEG KISS [17]; Myopoint [18]; EMOTIV[9]; EOG WEPU [19]; AlterEgo[10]) on accessibility and new technologies. The framework improves our knowledge and understanding on how disabilities are impacting artists abilities to engage with the Metaverse. The framework in Table 1 presents technical solutions listed in the reviewed literature that empower physically disabled artists to overcome the limitations of their specific disabilities. The framework helps foster interdisciplinary examination involving engineering, computer science, and creative and performing art to develop an understanding on how the Metaverse can help physically disabled users and content creators. The framework in Table 1 provides an indicative list of key technologies, with associated examples of collaborative activities, which need to be advanced and further developed in the Metaverse and provide guidance on accessibility in the Metaverse.

*Table 1:Framework of technologies and tools for enabling physically disabled artists to engage in the Metaverse concept.*

[Insert here]

The collaborative development activities outlined in the framework detailed in Table 1, place considerable importance on developing the complementary skills needed to empower physically disabled artists to (1) engage with activities related to creative and performing artist in the Metaverse, and (2) engage with the digital society in the Metaverse, contributing to a fair society and promoting high social skills and cultural awareness. In addition, activities for ensuring networking opportunities for disabled artists need to be included with knowledge transfer, entrepreneurship, and commercialisation networks with the Digital Research in the Humanities and Arts (DRHA) based groups, the International Conference on Social Science, Arts and Humanities (ICSAH), Arts & Disability Forum networking events, and active outreach activities, engagement, participation in (scientific) workshops, seminars, and conferences.

### 4.1. Increasing Awareness on Accessibility of the Metaverse

Although there are many potential commercialisation opportunities for creative and performing arts in the Metaverse (e.g., dance training in VR, music as NFTs), there is no intellectual property associated with this framework. The new framework can be freely disseminated to academic and Metaverse organisations, encouraging creative and performing arts in the Metaverse.

Building upon the framework of technologies and tools for supporting physically disabled artists in the Metaverse, Table 2 outlines the Metaverse economy, categorised by Market Cap, Market Cap Rank, Tokenomics (Token Price in USD) – data compared from two sources, (1) CoinMarketCap[11] and (2) CoinGecko[12] (as of 30th August 2022). Although big companies (e.g., Meta) are considered as the main future players in the Metaverse, the current top 10 Metaverse projects in 2022 (by market cap) are identified in Table 2.

---

[5] https://en.wikipedia.org/wiki/StyleGAN

[6] https://www.tiltbrush.com/

[7] https://en.wikipedia.org/wiki/DeepArt

[8] https://sougwen.com/

[9] https://www.emotiv.com/

[10] https://www.media.mit.edu/projects/alterego/overview/

[11] https://coinmarketcap.com/

[12] https://www.coingecko.com/

*Table 2: Summary of the top 10 projects (in 2022) of the Blockchain Metaverse economy.*

[Insert here]

Table 2 highlights that in 2022Ethereum-based (ERC-20) crypto tokens dominate the Metaverse economy. From the top 10 projects analysed (by Market Cap), only one project uses Bitcoin, the remaining 9 projects use ERC-20. Some of the top 10 Metaverse projects are directly related to arts, such as the RNDR project, designed for a 'distributed GPU rendering network built on top of the Ethereum blockchain, aiming to connect artists and studios in need of GPU compute power with mining partners willing to rent their GPU capabilities out.'[13] Other projects are predominately for play-to-earn gaming. However, the projects listed in Table 2 are expanding into NFTs, Decentralised Finance (DeFi), Decentralised Apps (Dapps), and other options in the Metaverse economy. Since the top Metaverse projects constantly change, the Table 2 represents a 'snapshot in time' of the Metaverse—from the lenses of 2022.

### 4.2. Blockchain Metaverse economy and usability by differently abled users

The projects summarised in Table 2, outline a very exciting new potential for online interaction, creativity, and economic participation, that have arisen with the development of the Blockchain Metaverse economy. In this study, we are interested in how accessible and inclusive these technologies are for Metaverse content creators and users with disabilities. A key component of creating a diverse and inclusive digital ecosystem is ensuring equitable opportunities and experiences for people with impairments.

Unfortunately, there is no information on the usability of these platforms by people with disabilities. There is no information provided on this topic in the webpages, in the white papers, or anywhere on the internet on the top 10 initiatives in the Blockchain Metaverse ecosystem – listed in Table 2. Nevertheless, the significance of accessibility issues in the creation and implementation of Metaverse technologies must be included, and this is the point we emphasise in this analysis.

The term 'accessibility' describes how technologies are created and put into use so that people with disabilities can access, use, and engage with digital content. Accessibility becomes essential in the Metaverse to enable people with disabilities to fully engage, contribute to the creation of content, and take advantage of the immersive experiences provided by these platforms.

In Figure 1 we list the most important factors than need to be considered while encouraging accessibility and inclusivity within the Metaverse ecosystem.

---

[13] https://coinmarketcap.com/currencies/render-token/

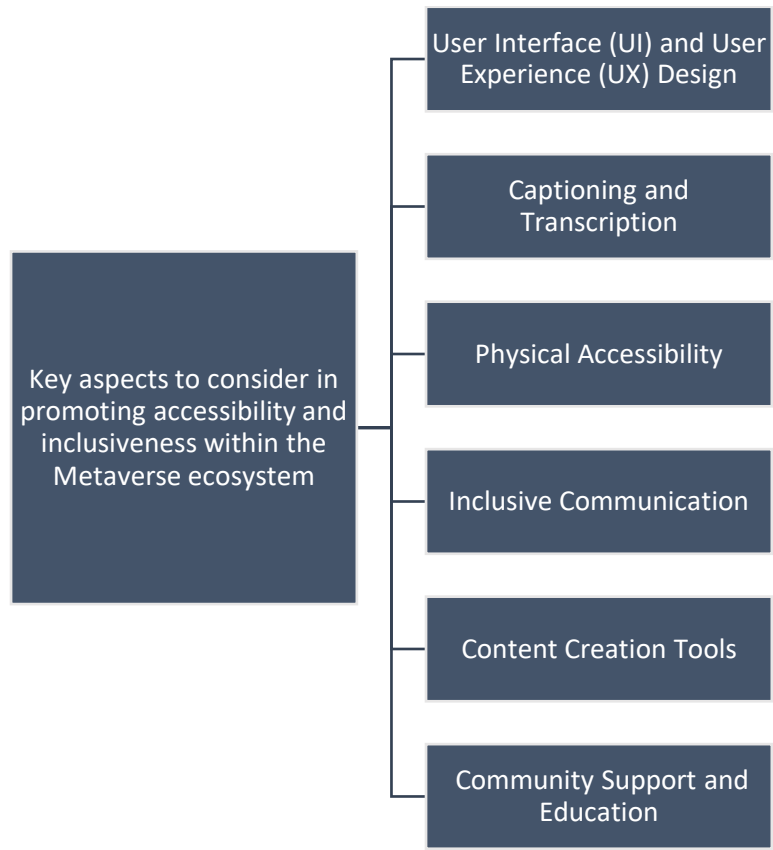

*Figure 1: Accessibility and Inclusivity in the Metaverse ecosystem*

In the following text, we will analyse the key concepts described in Figure 1.

Metaverse platforms should adopt inclusive design principles, considering elements like colour contrast, font size, alternative text for images, keyboard navigation, and compatibility with assistive technologies like screen readers and voice commands. Giving users customised UI options can improve accessibility even more.

Including closed captioning and transcription services for audio and video content within the Metaverse can guarantee equal access to knowledge and communication for those with hearing impairments.

Virtual reality (VR) or augmented reality (AR) devices may be used frequently in the metaverse. To ensure that people with mobility impairments can use these devices comfortably, it is essential that we consider the physical accessibility needs that need to be included in new devices.

People with speech impairments or other communication difficulties can benefit from the Metaverse's inclusion of accessible communication channels, such as text-based chat or speech-to-text capability.

Metaverse platforms should include tools for content creation that are flexible to users of various skill levels and talents. Disabled content creators can express themselves and contribute to the Metaverse economy with the help of intuitive interfaces, automation tools, and compatibility with assistive devices.

Support from the community and education about accessibility best practises can help to promote an inclusive environment and persuade designers, artists, and users to give accessibility considerations priority.

It is crucial to emphasise that inclusivity extends beyond technical accessibility features. A more inclusive and supporting digital environment can be produced by addressing social barriers, encouraging a sense of belonging, and actively incorporating disabled people in the Metaverse development processes.

Collaboration between developers, disability activists, and relevant stakeholders is also essential for new information and communication technologies to be inclusive and accessible in the Metaverse. The Blockchain Metaverse ecosystem can work towards a more accessible and inclusive future by incorporating accessibility concerns from the very beginning of development and regularly soliciting feedback from users with disabilities and content creators.

## 5. Output-Summary of the Review

The summary of the review was focused on identifying obstacles and solutions for supporting physically disabled artists (e.g., musicians, performers) to engage with the Metaverse, from the comfort of their home. The summary was performed with a group of experts from the University of Oxford (focus group, workshop, and interviews), on furthering opportunities for getting disabled artists involved in social and community events. The interviews were structured with open questions on the increased adoption of new technologies, including the Metaverse, in virtual realities and in everyday lives. Specific focus of the interviews was placed on the needs of physically disabled users, and the key new technologies that can help physically disabled people to operate in our society (e.g., work and socialise in hybrid physical, virtual, and augmented realities). The key categories of the Metaverse values for physically disabled creatives are outlined in Table 3.

*Table 3: Summary of Metaverse values for physically disabled creatives (and arts and humanities in general).*

[Insert here]

The framework of Metaverse values for physically disabled creatives in Table 3 defines a new interoperable digital economy, centred around portable digital assets or NFTs. This interoperability is however a very complex task. The complexity emerges firstly because such interoperability is not profitable or desirable for individual projects; and secondly because the computers at present lack the required computing power. For example, the Meta's Horizon Worlds can be used with maximum of 20 participants, and the Fortnite Battlefield 2042 is maxed in the range of 100 to 128 participants.

The design was grounded on a single intrinsic case study—based on Robert Yin [14]. GT was applied to categorise findings—coding, memoing, serendipity patterns, sorting, and writing, in combination with content and narrative analysis, and integrated with selective re-purpose open data from Metaverse platforms. Specific focus of the interviews was placed on the needs of physically disabled users, the key new technologies and the interoperability and limitations of the current Metaverse concept. The key findings are presents as anonymised statements in a summary table

*Table 4: Summary table of the key findings from interviewing experts in the field (statements are anonymised).*

[Insert here]

Around 20 experts participated in the interviews and workshops, their input was recorded in the summary table, around half of them were women, two experts were from disability groups. The main objective of the interviews outlined in the summary table in Table 4, was to configure a new framework for enabling physically disabled creative to engage with the Metaverse, via the associated physical and cognitive technologies and devices. To derive with the summary table, the research applied GT to categorise the statements (Table 4) and to identify the correct tools and technologies that would empower users with physical and cognitive disabilities—e.g., enabling Metaverses to allow users to operate haptic gloves to touch and feel art masterpieces on display (Table 1).

5.1. Analysis of the Metaverse is an Internet virtual world in the context of Web Content Accessibility Guidelines (WCAG) 2.2

This section emphasises the importance considering WCAG 2.2 compliance in the Metaverse context, because the Metaverse is an online virtual environment. The EU, UK, and US, have put in place legal frameworks to encourage accessibility and inclusivity, but WCAG 2.2 is especially relevant for accessibility and inclusiveness of disabled users and content creators.

The Web Accessibility Directive (2016/2102) in the European Union outlines the accessibility criteria for websites and mobile applications in the public sector. It encourages member states to implement accessibility measures in various digital services, including virtual environments, and is compliant with WCAG 2.1 Level AA. Adherence to the WCAG recommendations, especially version 2.2, would show the EU's commitment to accessibility.

The Equality Act 2010 in the UK prohibits discrimination based on a person's disability and mandates that service providers make reasonable modifications to guarantee equal access. The WCAG 2.1 Level AA standard for accessibility has been embraced by the UK government as well. It would be easier to fulfil legal requirements and improve accessibility for impaired users and content providers in the Metaverse by ensuring compliance with WCAG 2.2.

The Americans with Disabilities Act (ADA) of 1990, which mandates reasonable accommodations for people with disabilities, is applicable to virtual platforms in the United States. Although there is no formal federal statute requiring WCAG compliance, the ADA's duties have been interpreted in a number of court cases to include digital accessibility. The de facto norm is WCAG 2.0 Level AA, and adoption of the most recent version, WCAG 2.2, would show a commitment to accessibility and lessen the likelihood of legal issues.

Integrating WCAG 2.2 compliance into the creation and continuing management of the Metaverse is essential for the legal requirements in the EU, UK, and US. It would guarantee that people with disabilities may fully engage in virtual experiences and give content producers the tools they need to create inclusive and accessible content. Additionally, adhering to these rules and recommendations supports an inclusive culture and equal chances for everyone, regardless of disability, in the quickly changing environment of modern information and communication technology.

# 6. Results

We summarise the key findings in a new framework for supporting, enabling, and encouraging physically disabled creative and performing artists to participate and build creative communities based on shared values. The framework is targeted at configuring a Metaverse that can be accessible/usable/inclusive for people with disabilities—via the associated physical and cognitive technologies and devices.

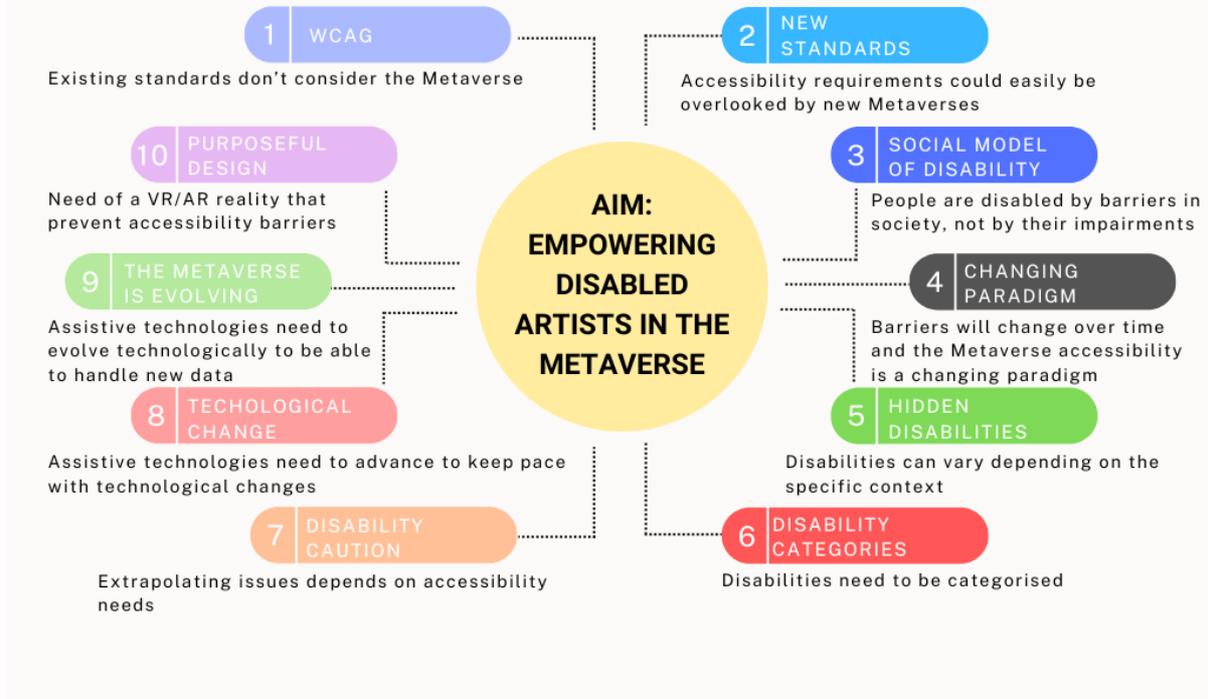

*Figure 2: New framework (FMFI) for enabling disabled creative and performing artists.*

The new framework in Figure 2 is designed based on a) state-of-the-art literature review, b) interviews of leading experts in the field, which included also c) physically disabled users. The new framework is guided by the need to build a purpose-driven VR/AR reality with 'purposeful design' [1], specifically targeted at supporting physically disabled artists by enabling and encouraging them to engage in the Metaverse projects. The framework identifies key 'barriers' disabled artists face that are preventing them from engaging with the Metaverse technology and describes the need to build flexibility to mitigate social barriers, including how 'disabled' creatives are perceived in the Metaverse, and consider whether in a virtual world artists can still (and want to) identify as disabled.

The mapping of future Metaverse requirements for inclusiveness and accessibility in Figure 3, is a first attempt to utilise the framework and develop appropriate standards. These include: design and introduction of knowledge exchange drivers that results in new standards, new social model of disability, changing paradigms, and new experiences that empower physically disabled artists through appropriate trends in user-led decentralised content, interactive narrative, and virtual personalisation strengthening the capacity of arts and humanities research to contribute profitably to disabled artists through the Metaverse context.

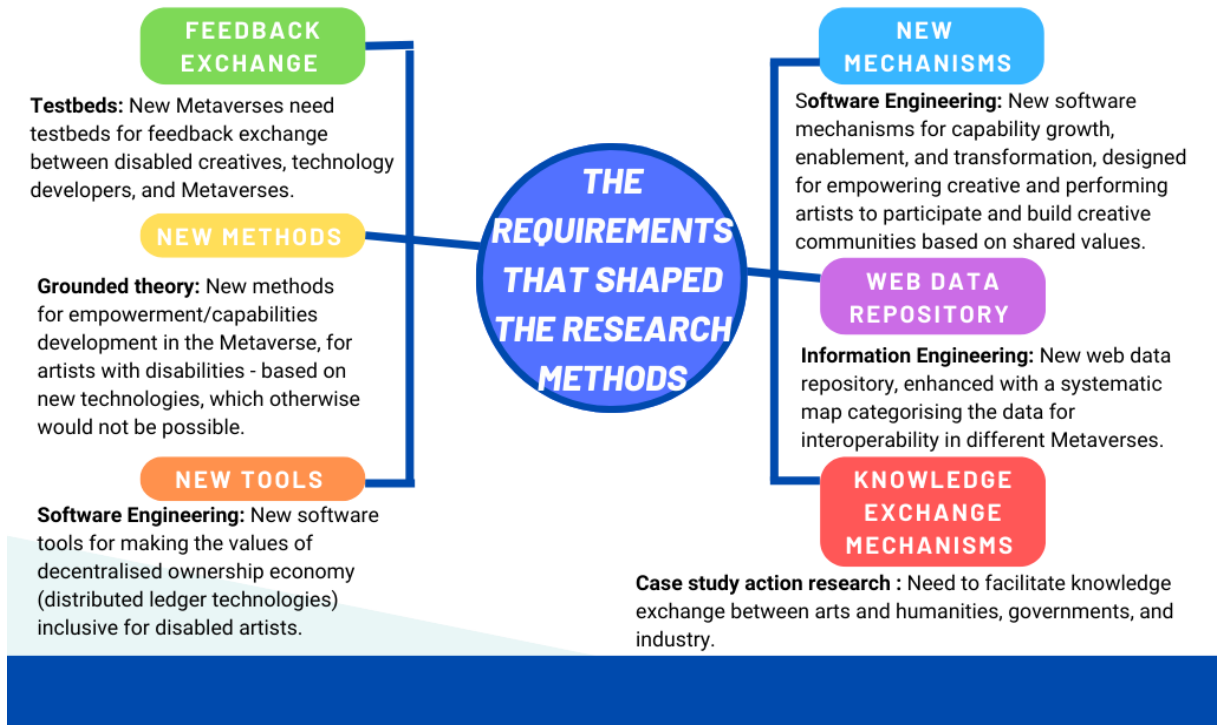

*Figure 3: Mapping the future Metaverse.*

The mapping in Figure 3 applies McCarthy and Wright's [20] deeper, culturally driven approach. It highlights that, through processes of citizen empowerment, the Metaverse will elevate content production from disabled artists up to the level of valued social experience. To facilitate this, the mapping refers to new methods for management and transactions, to store and exchange value and to create a positive impact on disabled artists participating in the governance of these projects/platforms. Figure 3 maps the overarching key philosophical, legal, and ethical challenges associated with the responsibility for making the future Metaverse inclusive of people with disabilities present in the academic literature.

6.1. Assessment and analysis: feasibility of the proposed mapping of the future Metaverse

The suggested paradigm for accessibility and inclusivity in the future Metaverse strives to empower artists who are physically challenged and remove the obstacles they encounter. The framework emphasises the necessity to reduce social barriers in the virtual environment and emphasises the need of intentional design. It also raises important considerations about how disabled artists are perceived and recognised in the Metaverse.

The mapping of future Metaverse accessibility and inclusivity needs shown in Figure 3 is developed to be used as the starting point for creating the required standards. The framework is a forward-thinking strategy for promoting inclusivity and may be seen in the combination of information exchange drivers and the exploration of a new social model of disability. The mapping aims to provide support experiences for artists who are physically disadvantaged by embracing shifting paradigms and utilising user-led decentralised material, interactive storytelling, and virtual personalisation.

A key component of the framework is the emphasis on improving how arts and humanities research may support disabled artists through the Metaverse. It acknowledges the Metaverse's potential to offer disabled artists a platform to flourish creatively and presents chances for multidisciplinary collaboration and creativity.

It is crucial to recognise, nonetheless, that putting this framework into practise and attaining its intended objectives may present difficulties. Collaboration amongst interested parties is necessary for the creation of acceptable standards and the adoption of a new social model of disability, including engineers, designers, disability advocates, and the artists themselves. To make sure that the standards and experiences are inclusive, accessible, and in line with the needs and goals of disabled artists, it is necessary to have ongoing discussions, conduct research, and make iterative adjustments.

The future frameworks should also consider any potential restrictions and moral questions related to virtual personalisation. Virtual personalisation can improve user experiences, but it's important to balance this with averting the construction of virtual identities that might be stigmatising or excluding.

Overall, the suggested framework mapping is designed to making progress towards removing obstacles for artists who are physically challenged and encouraging inclusivity in the Metaverse of the future. The fulfilment of an inclusive and supporting virtual environment for disabled artists will depend on collaborative efforts, ongoing research, and the engagement of diverse stakeholders. It serves as a foundation for further investigation and the development of appropriate standards.

### 6.2. Illustrations

To illustrate how accessibility and inclusiveness could operate in new information and communication technologies for disabled users and content creators in the Metaverse, we include some visualisations from the XR technologies currently used (in 2022).

In Figure 4 we illustrate an example of drawing (artistic creation) using MS Hololens 2 in one of the applications developed by PhishAR (a new start-up by a University of Oxford postdoc). In this case, the user simply pinches their two fingers to start drawing lines and objects in their space.

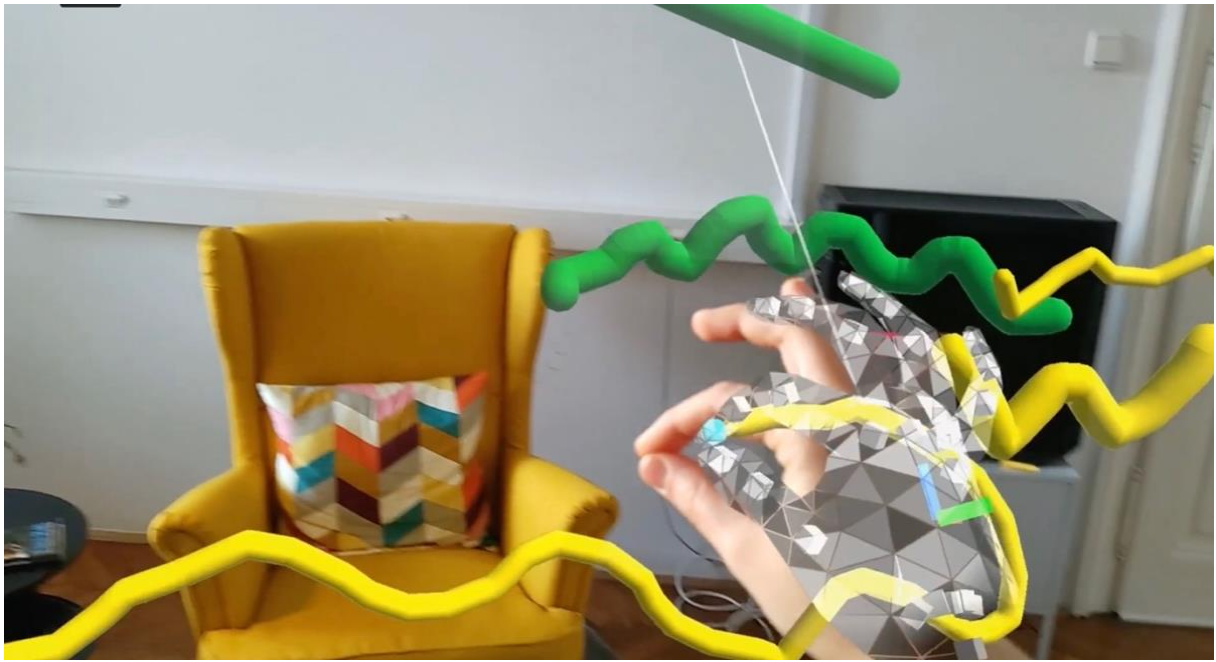

*Figure 4: Example of drawing (artistic creation) using MS Hololens 2.*

Figure 5 further showcases other ways of hand-based user interactions in MS HoloLens, and similarly in VR headsets: Users can manipulate and resize objects by pinching them and moving, press buttons (like piano keys), move sliders, etc. The artistic work of PhishAR exhibits a unique focus on intuitive and immersive experiences when used in conjunction with hand-based user engagements in MS HoloLens and VR headsets. The virtual environment gains a degree of realism and interactivity when users can pinch and move things to resize and change their shape. This type of gesture-based interaction blurs the boundaries between the real and virtual worlds by giving consumers a simple and straightforward way to interact with digital content.

Additionally, adding sliders and button presses like piano keys raises the level of user involvement and control. PhishAR gives users a sense of comfort and control in the virtual environment by simulating common real-life interactions. This not only improves the user experience but also makes it easier to interact and navigate through the artistic work.

The MS HoloLens and VR headsets' usage of hand-based user interactions demonstrates how these technologies may be used to create immersive and engaging experiences. PhishAR has taken advantage of these devices' capabilities to harness the power of human gestures and movements, allowing users to actively participate in the artistic creation. Users may now shape and customise their virtual experiences in a more tactile and engaging way, opening up new avenues for artistic expression and creative inquiry.

Overall, PhishAR's incorporation of hand-based user interactions in MS HoloLens and VR headsets is an example of how virtual reality and augmented reality technology are constantly improving. PhishAR paves the path for more immersive and engaging artistic experiences by allowing intuitive and interactive ways of manipulating objects, hitting buttons, and adjusting sliders, giving users a compelling and dynamic medium for creative expression.

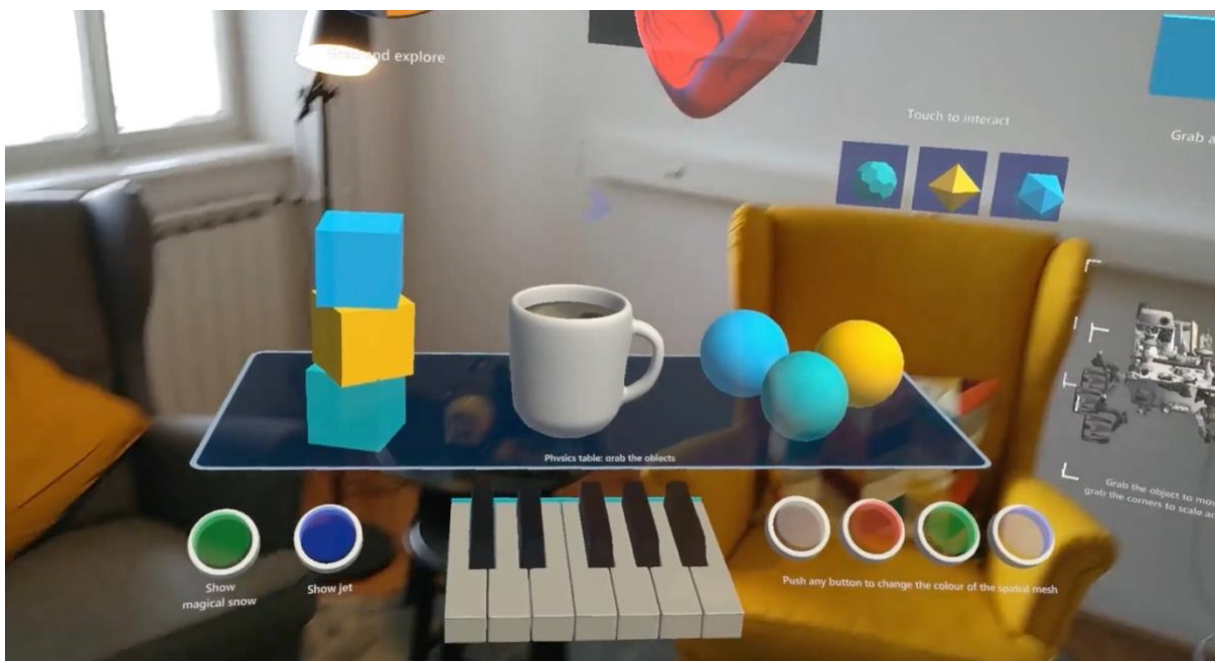

*Figure 5: Hand-based user interactions.*

Figure 6 is an example of actual looks of one the today's metaverse platforms ([Spatial.io](Spatial.io)).

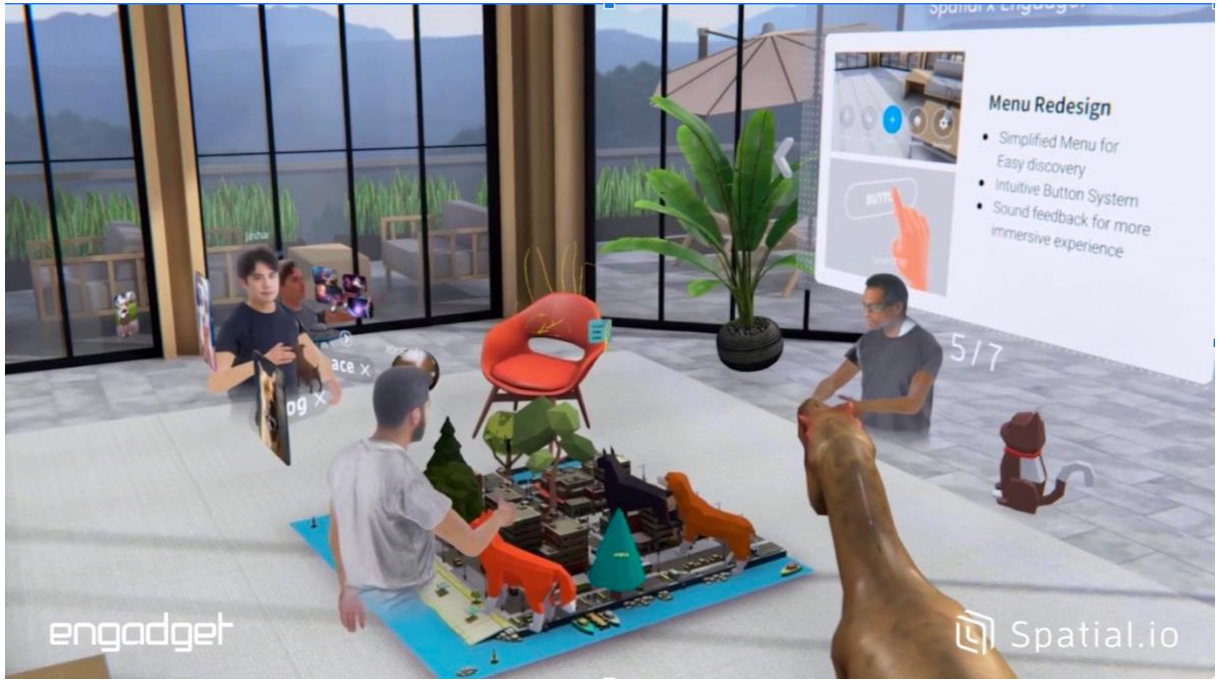

*Figure 6: Metaverse platforms.*

Figure 7 is an example of using Metaverse for 'serious' work, in which the two screens show the same situation from two different AR user's worldviews while they analyse multidimensional dataset in AR.

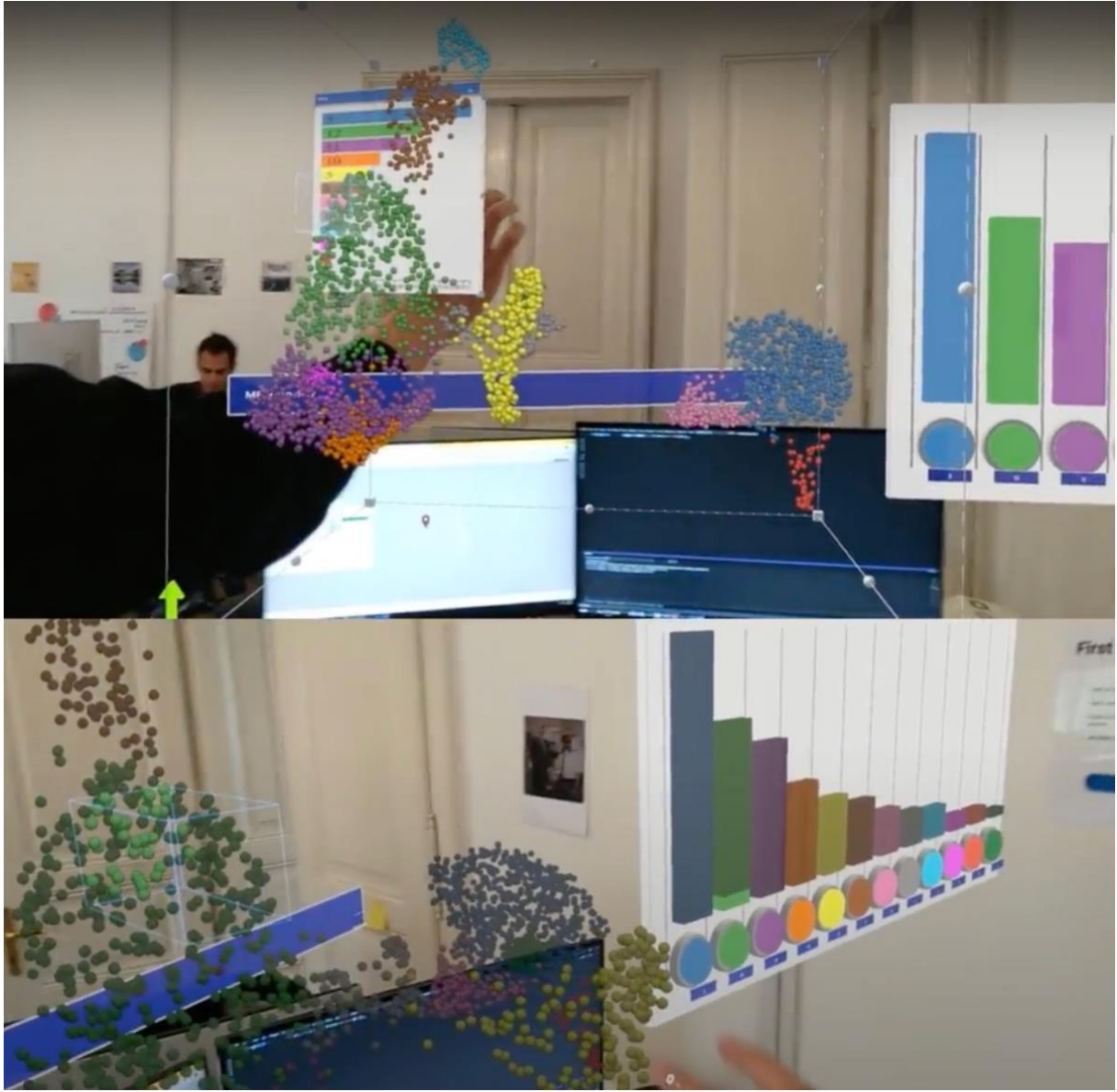

*Figure 7: Using Metaverse for work.*

The Metaverse is a virtual environment where people may connect, create, and explore, and it has emerged as a result of the rapid growth of technology. It is vital to keep this immersive digital environment open to and inclusive of users of all abilities as it develops popularity. With the illustration in Figures 4-7, we examined and assessed several facets of inclusion and accessibility in the Metaverse. In next sections of this article, we examine the use of universal design principles, the participation of disabled users in decentralised Metaverse projects, the function of governing bodies and regulatory agencies, a comparison of accessibility features among key players, and the effects of the digital divide on inclusivity.

6.3. Final examination and analysis of the key requirements for enhancing accessibility and inclusivity in the Metaverse

For the Metaverse to function, the adoption of universal design principles are necessary, especially because users in the Metaverse have a variety of demands and abilities. The goal of universal design is to make systems, environments, and products that can be used, understood, and accessed by people of all abilities. Developers of the Metaverse should think about include features like customised interfaces, alternate input methods, and support for assistive technologies in order to guarantee equal accessibility for all users. In addition, giving users with disabilities clear and explicit directions, simple navigation, and scaled information can greatly improve their user experience.

Decentralised Metaverse projects should actively include disabled users in the design and evaluation of their platforms to promote inclusivity. Engaging people with disabilities as advisors, forming teams specifically focused on accessibility, and doing user testing with people with disabilities are examples of practical ways to accomplish this. Developers can spot possible obstacles, hone user interfaces, and apply inclusive design principles by taking into account the thoughts and views of impaired users. Collaboration can be further facilitated by frequent feedback loops and inclusive design sprints, which can also make sure that the various demands of disabled users are properly met.

The accessibility and inclusivity of the Metaverse can be greatly enhanced by the involvement of governments and regulatory organisations. They can create and implement rules requiring accessibility requirements for online services. For developers designing accessible Metaverse experiences, guidelines already in place like the Web Content Accessibility Guidelines (WCAG) can be used as a guide. Providing financing or tax advantages to Metaverse projects that meet or exceed accessibility criteria is another way for governments to encourage them to prioritise accessibility. In addition, joint initiatives by governments, organisations that advocate for the rights of people with disabilities, and technology firms may result in the creation of thorough accessibility standards that are tailored to the Metaverse.

Major Metaverse player Meta has made progress integrating accessibility features into its platform. Meta has incorporated features like voice control, closed captioning, and changeable text sizes with a focus on diversity. To get input and enhance accessibility, they have also worked together with organisations for people with disabilities. To assure continuous improvement and responsiveness to user needs, however, persistent efforts are required. To develop more inclusive experiences for all users, other significant participants in the Metaverse market should aim to imitate and improve on Meta's accessibility initiatives.

For people with impairments in particular, the digital gap presents serious obstacles to inclusivity in the Metaverse. Accessibility covers the essential hardware and software for participation in addition to the software and platforms themselves. Barriers for people with disabilities might be caused by the cost of procuring accessible hardware, such as virtual reality headsets or specialised input devices. Technology firms, governments, and advocacy organisations must work together to provide inexpensive solutions, promote subsidy programmes, and make sure that everyone, regardless of financial situation, has access to and can use assistive technologies.

This examination was targeted at analysis of enhancing accessibility and inclusivity in the Metaverse, as depicted in the Figure 8. The diagram illustrates the interconnectivity of various aspects, including the application of universal design principles, involvement of disabled users in decentralised Metaverse projects, the role of governments and regulatory bodies, comparison of accessibility features among major players, and the implications of the digital divide on inclusivity. By exploring these interconnected factors, we can gain a holistic understanding of the measures needed to create a more accessible and inclusive Metaverse for all users.

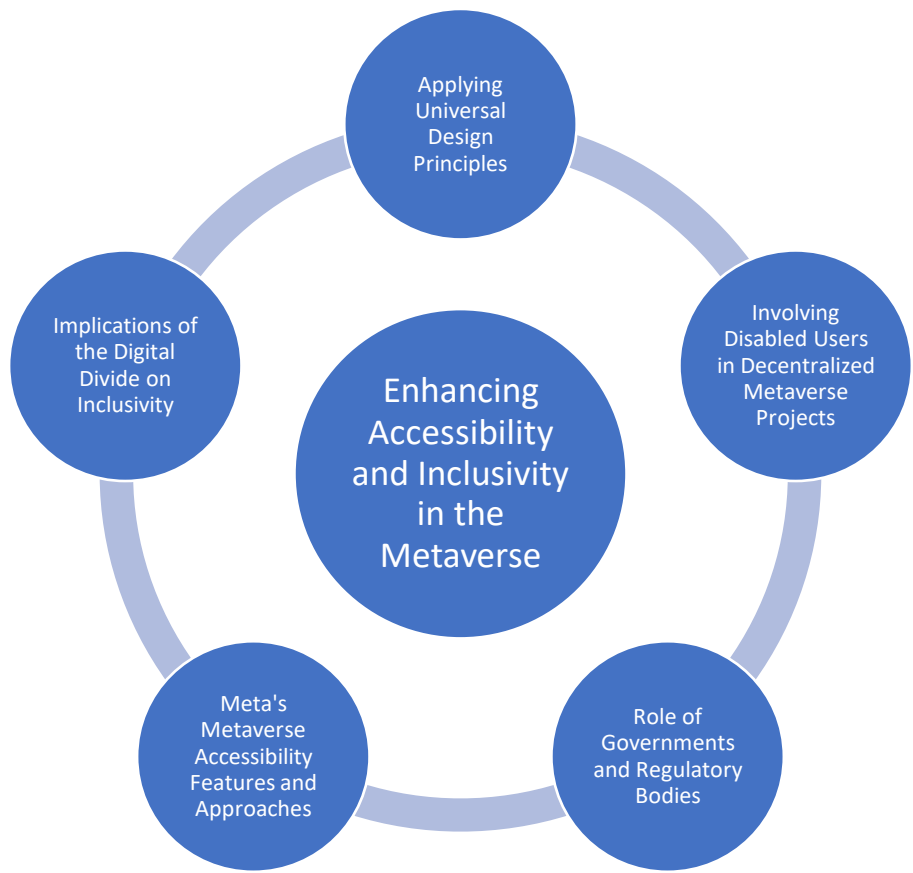

*Figure 8: Framework for unleashing the power of inclusion and crafting an equitable metaverse experience.*

The potential for innovation, creativity, and teamwork in the Metaverse is enormous. But for the Metaverse to be inclusive and accessible, everyone involved must work together. We can establish a digital environment that offers equitable opportunities and experiences for all users by adopting universal design principles, including disabled people in development processes, passing suitable rules, and tackling the digital divide. A more lively, diverse, and egalitarian digital future will result from the Metaverse's ongoing progress on accessibility and inclusivity.

## 7. Discussion

The key findings from this study are targeted at supporting disabled artists, to engage with the Metaverse. Although the findings can be transferable to creative and performing artists in general, the specific use cases that are considered in this study included the Metaverse technologies that enable (1) disabled users and content creators (artists) to advertise their products and services (e.g., dance or theatre performance), and (2) disabled users to test the service—in a virtual environment. The new framework of technologies and tools for enabling and encouraging disabled artists in the Metaverse (Table 1), the summary of the top 10 projects (in 2022) of the Metaverse economy (Table 2), and the framework of Metaverse values for disabled creatives (Table 3), empower disabled artists to build their own digital twin in the Metaverse, and test how audiences and users interact with the physical and cognitive environment. Such digital twins can help expand access to the marketplace. This would also be of great value to artists and consumers in emerging and frontier economies, opening new business opportunities and networking solutions. In the design process, focus was placed on building a framework that would inspire disabled artists to engage with the Metaverse and influence the development and the evolution of the Metaverse into a fairer, disability- and gender neutral, virtual and augmented reality that can be used by non-technical users and content creators.

This area has received minimum attention in research, and it will impact the future developments of the Metaverse.

In addition, a more normative contribution to the early discourse about the Metaverse for physically disabled artists, besides the requirements listed above, the list of dystopias collected by Sartor [15] relevant for our analysis. While the participation of physically disabled persons in the Metaverse may be well aligned with their human rights to be part of (virtual and augmented) communities, such participations may bring about also risks (or 'nightmares' in Sartor's view).

From the list of nine nightmares, at least six are applicable to the Metaverse. **First**, *Orwell's nightmare* may enable deeper surveillance with chilling-effect on people's behaviours. While physically non-handicapped individuals may opt to migrate from a region with increased surveillance, physically disabled individuals face increased difficulties in doing so. **Second**, despite the potential of the Metaverse to empower physically handicapped individuals' to further exacerbate their sense of self-esteem, dignity, and autonomy, the dystopic *Kafka's nightmare* highlight that such efforts can also effectively hinder these. Moreover, the virtual infinite memory [16] of IoT devices connected to the Metaverse may extend the persecution and political repression timeframe, or link an offence committed in a virtual world to a real individual. **Third**, as highlighted earlier the Metaverse can enable the dismantling of previous stereotypes associated with physical and cognitive handicaps, however, *Huxley's nightmare* highlight that future novel discriminations may also appear linked or within the Metaverse, with new forms of societal divisions. **Fourth**, the Metaverse can facilitate a new platform for exacerbating free speech and a free exchange of thoughts for artists, described as *Bradbury's nightmare*. On the other hand, existing social networks struggle to contain and limit the increasing number of attacks on race, religion, sex, or sexual orientation [21–23]. **Fifth**, the Metaverse can effectively facilitate a change in social arrangements where, for example, physical contacts are disregarded as undesirable, leading to a dystopic *Asimov's nightmare*. Such change in social arrangements may exacerbate the inclusion of people with handicaps in real-world environments. **Sixth**, *Nozick's nightmare* with his Experience Machine can exemplify the Metaverse with capabilities to provide stimuli that substitute and surpass real experiences. Such experiences, for example, may facilitate within the Metaverse albeit artistic but also detested or illegal sexual expressions [17,18].

7.1. Assessment and analysis of the dystopias related to the Metaverse for physically disabled artists, with a focus on the increased difficulties faced by these individuals.

From this initial high-level normative analysis of the future Metaverse we highlighted the need for careful weighing of the benefits and challenges brought about the future Metaverse for the artistic and existential expression of physically disabled individuals. Together with the proposed framework, further in-depth investigations will be needed to establish desired inclusiveness of the Metaverse. Because in the context of exploring dystopias related to the Metaverse for physically disabled artists, it is crucial to delve deeper into the specific challenges and hardships they might encounter. While the Metaverse offers promising opportunities for inclusivity and empowerment, it is essential to acknowledge the potential pitfalls and increased difficulties that disabled artists might face within this virtual realm. Issues such as limited accessibility features, barriers to navigation and interaction, and potential discrimination or stigmatization need to be thoroughly examined to ensure a more comprehensive understanding of the potential challenges faced by physically disabled individuals. By providing a more detailed analysis of these complexities, we can better address and mitigate any adverse impacts on disabled artists, fostering a truly inclusive and supportive environment within the Metaverse.

7.2. Discussion on the implications of the Metaverse on free speech, particularly regarding attacks on race, religion, sex, or sexual orientation, expanding into a discussion about the potential impact on disabled individuals.

The introduction of the Metaverse opens up both intriguing potential and worrisome consequences for free speech, particularly in light of the development of prejudice and hate speech that targets people based on their ethnicity, religion, sex, or sexual orientation. While the Metaverse offers a stage for a range of opinions and sentiments, it also creates opportunities for hateful people to spread hate and dangerous ideals there. Such assaults may intensify already prevalent societal stereotypes and the marginalisation of weaker communities. Additionally, it is vital to include the potential effects on disabled people in the conversation because they may already experience a lot of obstacles and discrimination in their everyday lives.

The Metaverse has the power to either improve accessibility and inclusivity for people with disabilities or to further marginalise and exclude them. To protect the rights and dignity of everyone, including those with disabilities, it is crucial to ensure that the Metaverse is built with robust measures to resist hate speech and encourage inclusivity. To promote a secure and respectful online environment where free speech may coexist with the protection of vulnerable populations, this calls for effective moderation systems, content regulations, and community norms. To ensure disabled people's full involvement and engagement in the Metaverse, it will also be crucial to provide accessible features that cater to their needs.

By addressing these issues, the Metaverse can develop into a location that encourages free expression and upholds the values of equality, respect, and inclusivity for all people.

## 8. Conclusion

This article critically addresses a pervasive and global challenge of enabling disabled individuals to effectively participate in modern society by examining the intersection of new and emerging technologies and the imperative for inclusiveness in the Metaverse. The proliferation of advanced technologies, such as artificial intelligence (AI) and the Internet of Things (IoT), within the Metaverse has underscored the significance and urgency of this research. The overarching impact of this study extends beyond the confines of academia, aiming to foster a more inclusive and equitable Metaverse that aligns with anti-disability discrimination regulations. Moreover, the research holds secondary value in unlocking technological opportunities through the development of innovative and autonomous devices designed to cater to diverse needs. By presenting a novel framework for integrating new technologies into existing Metaverses, this study enhances accessibility and inclusivity within these virtual realms. It offers a fresh perspective on leveraging technological advancements to prevent disability discrimination and gain early insights into the specific requirements of disabled individuals. Furthermore, the study sheds light on normative constraints and emphasizes the need for ongoing reflection and careful consideration to avert potential dystopian futures for individuals with physical disabilities in relation to the Metaverse. Through this comprehensive exploration, the research not only contributes to scholarly discourse but also provides actionable insights for policymakers, technologists, and society at large to foster an inclusive and empowering digital landscape for disabled individuals.

This paper reviewed and explored a topic that is interesting in the context of accessibility and inclusiveness in the decentralised Metaverses. The key area of focus was on the mechanisms to produce art by people with disabilities, through the emerging concept of the Metaverse. The Metaverse is, of course, not a new concept, and is decades old - but the extent to which communities of users with disabilities can be make artistic contributions in the decentralised Metaverses is a worthy new area that requires more exploring. The review derived with three fundamental concerns, each of which are interconnected:

1. **Users.** People with disabilities appear to be presented in the current accessibility and inclusiveness guidance as a single group. This is far from the case; everyone has a unique set of needs and requirements, and are supported by an ecosystem of people, communities, and support mechanisms. Although the focus on accessibility and inclusiveness appears to be about people with disabilities, it feels as if the decentralised Metaverses are developed in such a way that it is primarily about the project, rather than the people that may be affected by the project. When designing Metaverses (that are inclusive of people with disabilities), the emphasis should always be about the people first, and technology always second. Also, there is a lack of emphasis in current Metaverse projects about the use of methods to understand user needs; and there is a rich literature in interaction design that is there to be drawn upon.

2. **Art.** Art projects in the decentralised Metaverse are written about in a very nebulous or generic way. One way to write about art is the effect that it may have on those who perceive it, and those who produce it. Art production amongst communities of practitioners can have a very positive and therapeutic effect. If new and emerging Metaverse projects are interested in this area, it would be useful to look to local arts collectives or community groups for further inspiration. Study participants and reviewers identified few disability arts practitioner groups, one of them being called Dyspla[14]. More needs to be made of what is done with the art that is produced, and how it might be showcased or shared. There needs to be more than just *"this is a Metaverse project where disabled people produce art"*, for example, who are the producers, and how will the art be used?

3. **Funding.** Given the extensive amount of money that Meta is investing into their Metaverse, it might be worth for new Metaverse project in the decentralised space to seek guidance from them, in addition to more traditional government regulations. Disabled users are users too, and they will have a vested interest in ensuring that whatever they produce is as accessible as possible. It would also look good from their perspective if they were working with a university, as well as community stakeholder groups. There are many different (potential) use cases for the Metaverse, and Arts is an interesting example of one of them. For accessibility and inclusiveness framework or standard to be successful, we need more direct and firm engagement from disabled artists. Whilst there is a clear intent from new Metaverse projects to be accessible and inclusive, in their whitepapers, there isn't any clear expression of clear intent to invest specific amounts of time; numbers always helps when it comes to clarity.

This review study highlighted that although the accessibility and inclusiveness of new information and communication technologies for disabled users and content creators in the Metaverse is considered by new decentralised Metaverse projects as an interesting and worthy area, much more emphasis on the people and community side is needed to create a stronger case that the Metaverse is supporting disabled people and artists.

### 8.1. Limitations and areas for further research

To test and verify the performance of the new framework, specific test scenarios can be constructing on existing testbeds. The testing scenarios can verify the practical capabilities and can be applied by different Metaverse for enabling disabled artists to engage productively with the society. This could also be applied for designing the new Metaverses to improve their ability to adapt to enabling users and content creators with different disabilities.

### 8.2. Personal motivations for this study

While the authors of this study acknowledge that their areas of expertise is in computer sciences and engineering, the research study is an attempt to build new knowledge by analysing the problem from a very different perspective. The expertise of the lead author are in cybersecurity, and the study is heavily influenced by the viewpoint of a data and privacy security professional. In year 2021/22, after a series of surgeries, the lead author was temporary disabled, unable to walk, talk, or eat. This brief experience with a disability, is the main motivation for working this article.

---

[14] https://dyspla.com